\documentclass[aps,prd,reprint,longbibliography,preprintnumbers,floatfix,superscriptaddress]{revtex4-2}
\usepackage{graphicx}
\usepackage{booktabs}
\usepackage{amsmath,amssymb}
\usepackage[hidelinks]{hyperref}

\newcommand{\nConfMaxDiff}{0.01}
\newcommand{\nAECloAllCIMarg}{0.504}
\newcommand{\nAECloAllCICond}{0.511}
\newcommand{\nAECloAllMuMarg}{0.502}
\newcommand{\nAECloAllMuCond}{0.503}
\newcommand{\nAECloAllVtxMarg}{0.504}
\newcommand{\nAECloAllVtxCond}{0.508}
\newcommand{\nAECloAllRecoilMarg}{0.502}
\newcommand{\nAECloAllRecoilCond}{0.503}
\newcommand{\nAECloAllN}{553{,}933}
\newcommand{\nAECloATwoPCIMarg}{0.497}
\newcommand{\nAECloATwoPCICond}{0.501}
\newcommand{\nAECloBTwoPCIMarg}{0.505}
\newcommand{\nAECloBTwoPCICond}{0.499}
\newcommand{\nCovTwoPLtHundredth}{30\%}
\newcommand{\nCovTwoPLtEqual}{49\%}
\newcommand{\nEnsPullAllBins}{114}
\newcommand{\nAECloATwoPRes}{0.004}
\newcommand{\nAECloAllRes}{0.001}

\newcommand{\fermilab}{Fermi National Accelerator Laboratory, Fermilab, P.O. Box 500, Batavia, Illinois 60510, USA}
\newcommand{\rochester}{Department of Physics and Astronomy, University of Rochester, Rochester, New York 14627, USA}

\begin{document}

\title{A Surrogate Detector Model for the MINERvA Open Dataset}
\preprint{FERMILAB-PUB-26-0789-CSAID-PPD-T}

\author{Sparshita Dey}
\affiliation{\fermilab}
\author{Alexander Himmel}
\affiliation{\fermilab}
\author{David Last}
\affiliation{\rochester}
\author{Liang Liu}
\affiliation{\fermilab}
\author{Pedro Machado}
\affiliation{\fermilab}
\author{Vishvas Pandey}
\affiliation{\fermilab}

\author{Linyan Wan}
\email{lwan@fnal.gov}
\affiliation{\fermilab}

\date{\today}

\begin{abstract}
The growing availability of open neutrino data promises broader scrutiny of interaction models, but a critical barrier remains: mapping theoretical predictions into reconstructed observables, which previously required computationally intensive methods inaccessible outside the original experimental collaborations. In this work, we present a probabilistic surrogate of the MINERvA detector response, trained on an open simulation dataset to map generator-level final-state particles and interaction geometry to reconstructed observables, without relying on generator-specific interaction labels. We assess its fidelity through conditional response distributions, reconstructed observables, and classifier-based closure tests. We further demonstrate generalization beyond the limitations of conventional phase-space-dependent reweighting by withholding two-particle–two-hole interactions entirely from training and showing that the surrogate reproduces their reconstructed response within the estimated uncertainties. In an end-to-end application to MINERvA charged-current inclusive muon-neutrino data, we reproduce selection efficiencies and purities and forward-fold predictions from GENIE, NuWro, and GiBUU for direct comparison with measurements. By moving theoretical predictions into reconstructed space rather than unfolding the data, this work establishes a reusable interface between neutrino interaction models and experimental observations, expanding the scope of physics accessible through open datasets.

\end{abstract}

\maketitle

\section{Introduction}

Rapid advances in artificial intelligence and machine learning have opened new avenues for extracting physics from existing high-energy physics (HEP) datasets, offering opportunities to revisit measurements and explore questions beyond the scope of their original analyses. These tools hold particular promise for overcoming some of the longstanding barriers to exploiting open experimental data. Yet interpreting HEP data requires more than access to the measurements themselves. Detector acceptance, finite resolution, imperfect particle identification, and the inherent ambiguity between detector observables and the underlying physical processes all complicate the extraction of physics information. Modern experiments account for these effects through high-fidelity simulations of the underlying interactions and the subsequent propagation of final-state particles through the detector, together with systematic uncertainties that characterize limitations in the modeling. Sophisticated reconstruction algorithms are then applied to the detector data to infer the properties of the underlying interactions, transforming complex detector signals into observables suitable for physics analysis. The resulting datasets can contain hundreds of reconstructed variables per event, linked by intricate, non-trivial correlations.

These complex software systems are often inaccessible to researchers outside the collaboration and even if it were possible, typically require substantial computing resources. Reweighting the released simulated events has been used for this purpose~\cite{Andreassen2020}: each event is given the ratio of the new model's prediction to the original one, so that the released detector response is carried over to the new model. Reweighting, however, only works where the released simulation already covers the phase space of the new model, and large weights dilute the statistical power of the sample~\cite{Kish1965}, as Section~\ref{sec:holdout} shows for two-particle--two-hole interactions. Where it does not, the detector response has to be interpolated or extrapolated from the events that exist, which is nontrivial in the high-dimensional space of the final-state particles~\cite{Hastie2009}. Surrogate detector models~\cite{Paganini2018,Amram2023}, trained on the released simulated events, are the right tool for this task: they learn the detector response as a function of the final state, can be applied to any generator's events with high accuracy, carry an uncertainty that can be validated, and run much faster than the full simulation and reconstruction. Using a surrogate model to fold theoretical predictions into the detector-observable space, the challenges of unfolding~\cite{DAgostini1995,Hocker1996,Koch2020} can be avoided, especially in novel variables and selections not previously studied by the experiment.

In this paper, we present the design, training, performance, and validation of such a surrogate model operating on the MINERvA open dataset~\cite{opendata}. The MINERvA detector~\cite{Aliaga2014} was a fine-grained tracking calorimeter located on-axis in the NuMI beamline at Fermilab. The detector was constructed from planes of triangular plastic scintillator strips (33~mm base, 17~mm height) in 3 orientations, featuring a dedicated upstream region of passive nuclear targets, a central fully-active tracking region, and surrounding electromagnetic and hadronic calorimeters. It utilized the magnetized MINOS near detector~\cite{Michael2008} immediately downstream as a muon spectrometer to measure momentum and charge. MINERvA recorded data in neutrino-enhanced (forward horn current, FHC) and antineutrino-enhanced (reverse horn current, RHC) horn configurations in both the Low-Energy (LE) and Medium-Energy (ME) target configurations from 2009 to 2019, with neutrino energy spectra peaking near 3.5\,GeV and 6\,GeV respectively~\cite{Aliaga2016,Valencia2019}. The collaboration has generously made its complete datasets public, including the recorded data, the corresponding simulation, the collaboration's analysis software and an example tutorial analysis~\cite{opendata}.

Section~\ref{sec:surrogates} describes the surrogate model and its design, training, performance, and uncertainty evaluation. Section~\ref{sec:holdout} validates the model on an interaction mode withheld from training. Section~\ref{sec:analysis} shows a MINERvA analysis with alternative generators as an example use case. Finally, Section~\ref{sec:conclusion} discusses the further possibilities opened up with these new tools.

\section{Surrogate Models}\label{sec:surrogates}

\subsection{Dataset}
\label{sec:dataset}

We use playlist 1A from the MINERvA open simulation dataset~\cite{opendata}, which provides the 15.8 million simulated neutrino interactions used in this work, 11.5 million of them charged-current $\nu_\mu$ interactions, in the ME FHC beam configuration with water and helium nuclear targets empty.
For each interaction, the dataset provides the generator-level truth record, together with the reconstructed event information when a muon candidate is identified.
The MINERvA collaboration produced this open simulation dataset using the GENIE neutrino event generator, version 2.12.6~\cite{Andreopoulos2010} (modified to add the Valencia two-particle--two-hole (2p2h) model~\cite{Schwehr2016}), a Geant4~\cite{Agostinelli2003} model of the MINERvA detector~\cite{Aliaga2014} and the downstream MINOS near detector~\cite{Michael2008} with beam-spill data overlaid for pile-up, and the MINERvA reconstruction framework.
Reproducing this complete experiment-specific chain outside the collaboration would require substantial software infrastructure and computational resources. The surrogate instead learns directly from the released simulation, without requiring any access to the original software.

Tables~\ref{tab:gen-vars} and~\ref{tab:reco-vars} list the variables used by the surrogate. Its input is restricted to what physically enters the detector: the final-state particles after intranuclear re-interactions, with the true vertex and the target nucleus as context.
Generator-specific interaction labels, together with quantities describing the underlying interaction such as the squared four-momentum transfer $Q^2$, the invariant mass of the hadronic system $W$ and the neutrino energy, are deliberately excluded, so the model can be applied to the final states of any generator, regardless of differences in how they may label interactions. Its output is the reconstructed event in the format of the open simulation dataset. The model surrogates every interaction in the open simulation dataset's training volume: charged and neutral current, the four neutrino flavors and all nuclei of the detector.

\begin{table}[tbp]\centering\footnotesize
\begin{tabular}{ll}
\toprule
Category & Variables \\
\midrule
Interaction & vertex position \\
 & target nucleus $(Z,A)$ \\
\addlinespace
Final-state particles & species \\
 & three-momentum \\
\bottomrule
\end{tabular}

\caption{Generator-level inputs of the surrogate.}
\label{tab:gen-vars}
\end{table}

\begin{table}[tbp]\centering\footnotesize
\begin{tabular}{ll}
\toprule
Category & Variables \\
\midrule
Muon & candidate found \\
 & MINOS match \\
 & charge sign \\
 & momentum and direction \\
\addlinespace
Vertex & reconstructed vertex position \\
\addlinespace
 Calorimetry & recoil energy \\
 & non-muon non-vertex energy \\
 & isolated-blob energy \\
\addlinespace
Hadron prongs & number of prongs \\
 & per prong: direction \\
 & per prong: pion- and proton-hypothesis momenta \\
 & per prong: proton score \\
 & per prong: exiting flag \\
\bottomrule
\end{tabular}

\caption{Reconstructed quantities predicted by the surrogate. The recoil energy is an estimator of the total non-lepton energy; the non-muon non-vertex energy uses the hits outside 100\,mm of the vertex; isolated blobs are localized groups of hits in at least two views.}
\label{tab:reco-vars}
\end{table}

\subsection{Model}
\label{sec:model}

\begin{figure*}[htbp]\centering
\includegraphics[width=\textwidth]{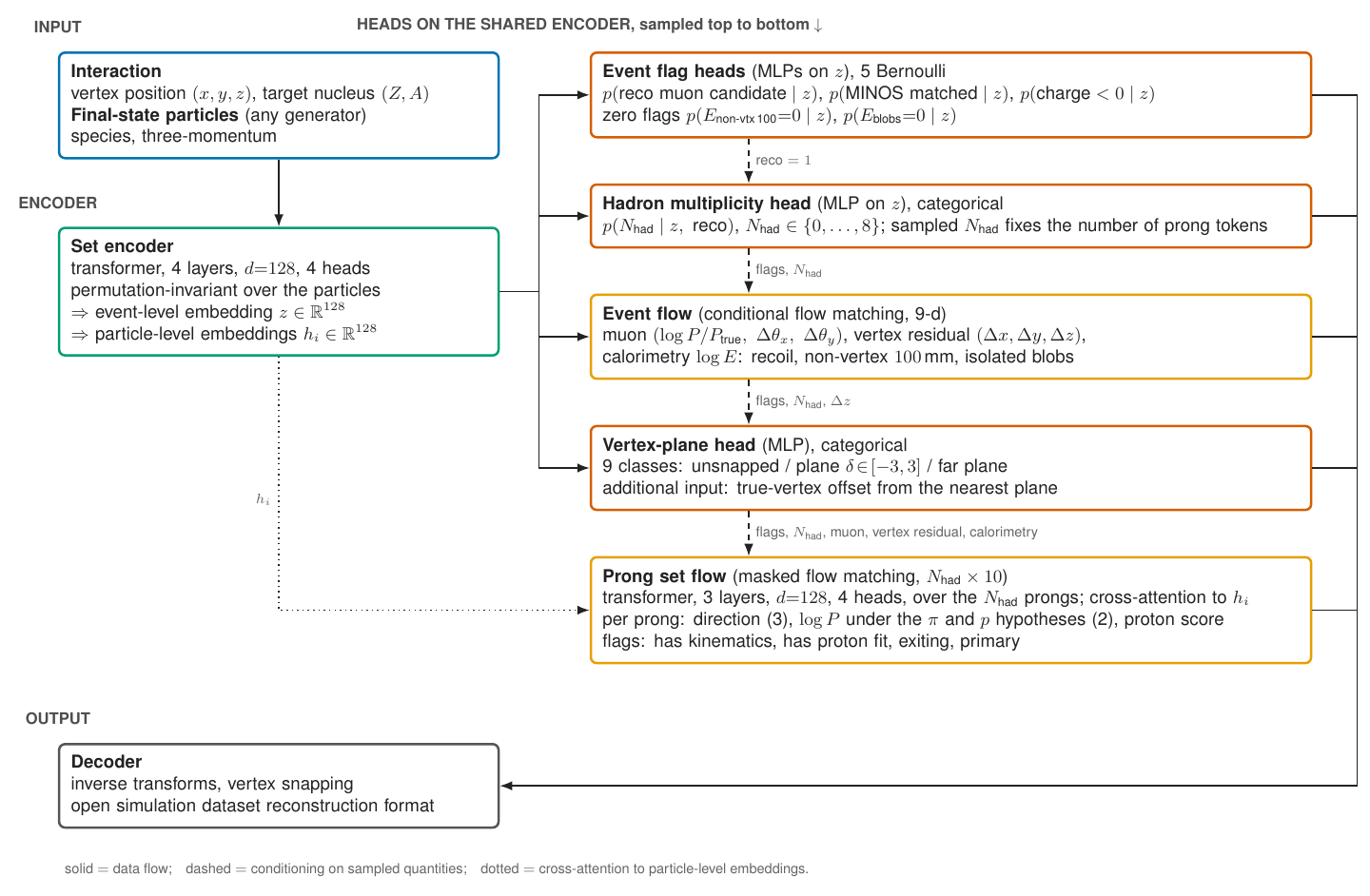}
\caption{Architecture of the surrogate. Solid arrows are data flow, dashed arrows conditioning on quantities
sampled by an earlier head, the dotted arrow cross-attention to the particle-level embeddings.}
\label{fig:schematic}
\end{figure*}

Figure~\ref{fig:schematic} shows the model. A transformer~\cite{Vaswani2017} set encoder~\cite{Zaheer2017,Lee2019} over the
final-state particles, with a global token~\cite{Devlin2019} carrying the interaction vertex position and target nucleus, produces two outputs: an event-level embedding used by the event-level heads, and particle-level embeddings, one vector per final-state particle, which the prong flow attends to; all other model outputs are produced by additional heads on these outputs, sampled hierarchically. Classification heads
predict whether a muon candidate is reconstructed, whether it is matched in MINOS, its charge sign, whether each of
the two non-vertex calorimetric energies is exactly zero, and the number of hadron prongs ($N_\mathrm{had}$). A conditional flow-matching model~\cite{Lipman2023} generates the event-level quantities: the muon momentum as a response to the true muon, the vertex residual and the three calorimetric energies, given the event flags and the $N_\mathrm{had}$ drawn by the classification heads. A vertex head then predicts whether the reconstructed vertex snaps to a scintillator plane, that is, whether the reconstruction places it at the plane position, as it does for most events, given the sampled vertex residual.
Finally a second set flow transformer that attends to the particle-level embeddings generates exactly $N_\mathrm{had}$ hadron prongs with their directions, their momenta under the pion and proton hypotheses, and their proton scores and flags.
A decoder writes the result in the reconstruction format of the open simulation dataset.

Hadrons with kinetic energy below 10\,MeV were dropped from the input during our training to improve computational efficiency and reduce training time: they contribute very little to the reconstructed variables, yet significantly worsen the computing performance. Neutrons above this threshold are retained, as their secondary interactions can contribute to the reconstructed calorimetric energy and hadron-prong multiplicity.

Events are split by subrun into a training sample of 12{,}640{,}819 events, 80\% of the 15{,}799{,}095 interactions of the open simulation dataset used in this work, a validation sample of 1{,}582{,}703 events, 10\%, and a test sample of 1{,}575{,}573 events, 10\%; all results shown are on the test sample. All heads are trained jointly with teacher forcing~\cite{Williams1989}, each head seeing the true values of the quantities it is conditioned on, so that the hierarchy is learned without sampling during training. At inference, the sampled outputs of each head instead provide the conditions for those that follow, allowing a complete reconstructed event to be generated. The flows use the conditional flow-matching objective~\cite{Lipman2023} and the classification heads use the cross-entropy, the negative log-likelihood of the observed class~\cite{Goodfellow2016}, as the objective. Training uses the AdamW optimizer~\cite{Loshchilov2019} with a one-cycle schedule~\cite{Smith2017} and proceeds in two stages, the encoder with the event-level heads first and the vertex head and prong flow added in a warm-started second stage.

\subsection{Performance}
\label{sec:performance}

We first evaluate the performance by comparing the reconstructed variable response as a function of true variable value between the open simulation dataset test sample and the surrogate. In bins of a true quantity, the median, fraction, or mean of a reconstructed quantity is compared between
the open simulation dataset with its statistical error and the surrogate output with its predicted uncertainty
(discussed in more detail in Section~\ref{sec:uncertainty}). The figures and tables in this section all use the full test sample.

\begin{figure*}[tbp]\centering
\includegraphics[width=0.95\textwidth]{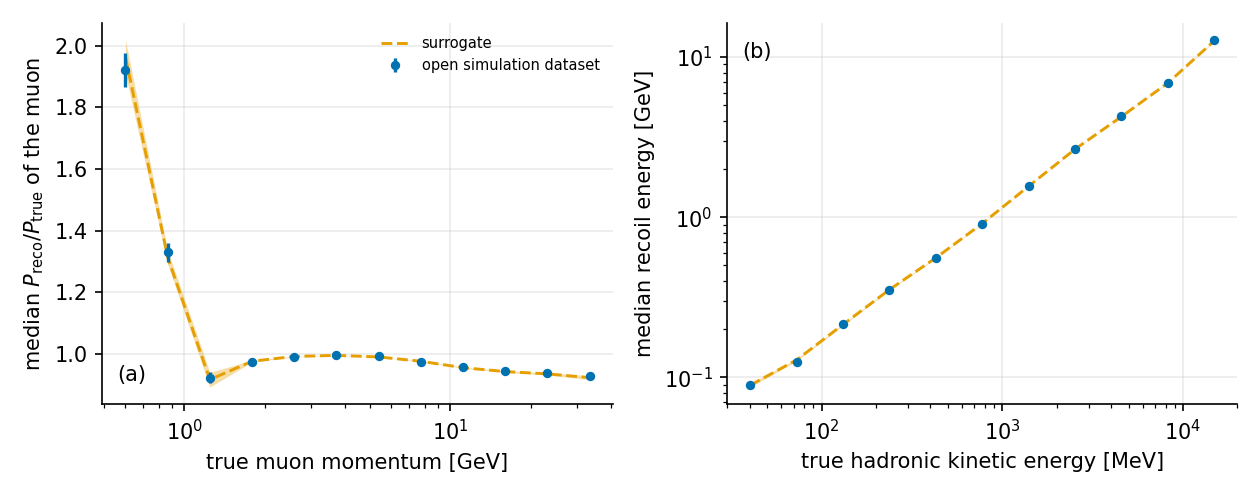}
\caption{(a) Median ratio of reconstructed to true muon momentum versus true muon momentum, for events with a true muon. (b) Median reconstructed recoil energy versus the true kinetic energy of the final state hadrons. Points: open simulation dataset with statistical errors. Lines and bands: surrogate, mean and spread over the four ensemble members, each sampled with the same random numbers (the model uncertainty of Section~\ref{sec:uncertainty}).}
\label{fig:response}
\end{figure*}

\begin{figure*}[tbp]\centering
\includegraphics[width=0.8\textwidth]{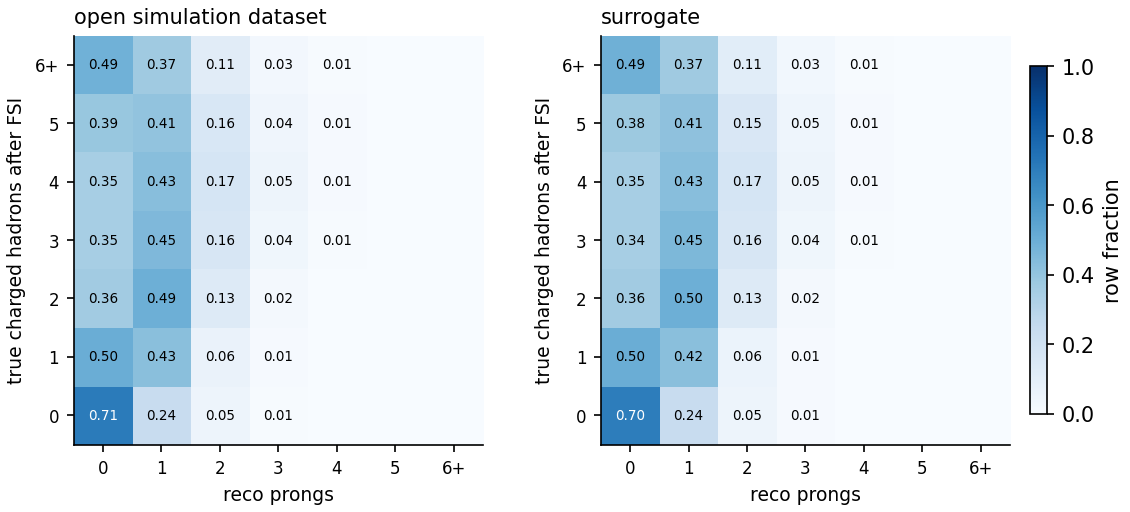}
\caption{Reconstructed prong multiplicity versus true charged hadrons, row-normalized, for the open simulation dataset
(left) and the surrogate (right). The largest difference in any two cells is 0.01, consistent with the statistical uncertainty of the cells.}
\label{fig:mult-conf}
\end{figure*}

Figure~\ref{fig:response} shows the muon momentum response and the recoil energy as functions of the truth: both follow the open simulation dataset over the full range within the predicted uncertainty. Figure~\ref{fig:mult-conf} shows the probability of each reconstructed prong multiplicity given the true number of charged hadrons, for the open simulation dataset and the surrogate; the two agree to \nConfMaxDiff{} in every cell, within the statistical uncertainty of the cells.

\begin{figure*}[tbp]\centering
\includegraphics[width=0.9\textwidth]{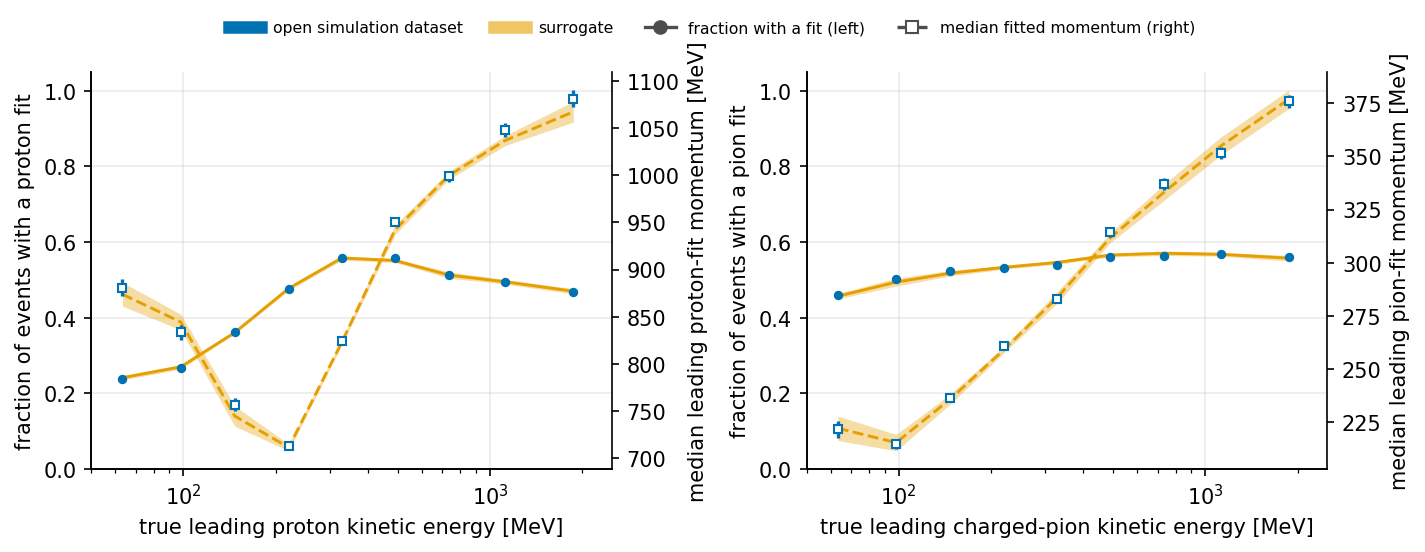}
\caption{Hadron reconstruction against the truth, for protons (left) and charged pions (right): fraction of events
with a fit under that hypothesis (circles and solid lines, left axis) and median leading fitted momentum (squares
and dashed lines, right axis) versus the true kinetic energy of the leading hadron of that species. Points: open simulation dataset with statistical
errors. Lines and bands: surrogate, mean and spread over the four ensemble members, each sampled with the same
random numbers.}
\label{fig:hadron}
\end{figure*}

As shown in Fig.~\ref{fig:hadron}, particle identification and the hadron kinematics are tested per species: the fraction of events with a fit under the proton or the pion hypothesis and the leading fitted momentum, versus the true kinetic energy of the leading proton or charged pion, agree with the open simulation dataset over the full range.

To assess the agreement beyond individual response distributions, we train a gradient-boosted classifier~\cite{Friedman2001} to distinguish events from the open simulation dataset from those generated by the surrogate~\cite{LopezPaz2017,Ke2017,Pedregosa2011}. The classifier is trained on half of the test sample and evaluated on the remaining half event by event. We then measure its discriminating power by the area under the receiver-operating-characteristic curve (AUC)~\cite{Fawcett2006}, the test established for generative detector surrogates in calorimeter simulation~\cite{Amram2023,Krause2025}. An AUC of 0.5 indicates that the classifier is unable to distinguish the two samples using the information available to it. We test this first using reconstructed quantities alone, and then introduce summary features of the true final state, asking the more stringent question of whether the surrogate reproduces the detector response conditional on the underlying interaction.

\begin{table}[bt]\centering\footnotesize
\setlength{\tabcolsep}{5pt}
\begin{tabular}{lc}
\toprule
Classifier input & test sample \\
\midrule
reco variables & \nAECloAllCIMarg{} \\
\quad muon block & \nAECloAllMuMarg{} \\
\quad vertex block & \nAECloAllVtxMarg{} \\
\quad recoil block & \nAECloAllRecoilMarg{} \\
reco variables + truth features & \nAECloAllCICond{} \\
\quad + muon block & \nAECloAllMuCond{} \\
\quad + vertex block & \nAECloAllVtxCond{} \\
\quad + recoil block & \nAECloAllRecoilCond{} \\
\midrule
test events & \nAECloAllN{} \\
\bottomrule
\end{tabular}

\caption{Closure classifiers on the seven event-level variables (muon momentum and direction, vertex, recoil energy): all seven and each block, without and with the truth features, on the test sample. 0.5 means indistinguishable from the open simulation dataset.}
\label{tab:closure}
\end{table}

Table~\ref{tab:closure} presents the results for seven event-level variables commonly used in analyses: the muon momentum and direction, the reconstructed vertex, and the recoil energy. The classifier achieves an AUC of \nAECloAllCIMarg{} using reconstructed quantities alone and \nAECloAllCICond{} when the truth features are included, where \nAECloAllRes{} is the statistical fluctuation of the AUC of an uninformative classifier on a sample of this size. This deviation is mainly driven by the vertex block. Analyses studying interactions in the fully-active tracking region, such as the one in Section~\ref{sec:analysis}, use the reconstructed vertex through broad fiducial-volume cuts, and there we demonstrate that the impact of the residual differences on efficiency and purity at fiducialization is negligible.

Importantly, the agreement sought here is not between individual reconstructed events, but between their distributions and conditional responses. Detector simulation is inherently stochastic: two realizations of the same final state need not produce identical reconstructed events, but should follow the same underlying response distribution.

\subsection{Uncertainty Treatment}\label{sec:uncertainty-treatment}
\label{sec:uncertainty}

The surrogate aims to reproduce the open simulation dataset, and we evaluate the potential deviation as the surrogate's own uncertainty. The systematic uncertainties of MINERvA simulation datasets, such as the detector model uncertainties, are not evaluated here. The surrogate's output carries two uncertainties. The first is its sampling error: the model is stochastic, like the underlying simulation itself, so a second draw from the same truth gives a different reconstructed event. This spread reflects the learned detector resolution, which the closure tests above show to be correct.
The second is the uncertainty in the learned detector response: while each model is trained to reproduce the same simulation, the response it learns depends in part on the particular weights reached during training and can vary between different training runs. We capture this variation using an ensemble of four independently trained models, identical in architecture and training data but differing in their random initialization and data-shuffling seeds \cite{Lakshminarayanan2017}. Together, these models provide four plausible realizations of the learned detector response, whose spread serves as an estimate of the surrogate's model uncertainty. By treating each realization as a systematic universe, following the convention used by MINERvA~\cite{Aliaga2016} and by the analysis tools of the open data product~\cite{opendata}, we can carry this uncertainty through an analysis while preserving its correlations between reconstructed bins; Section~\ref{sec:analysis} demonstrates this procedure. The ensemble captures differences arising between independent training runs.

\begin{figure}[b]\centering
\includegraphics[width=0.9\columnwidth]{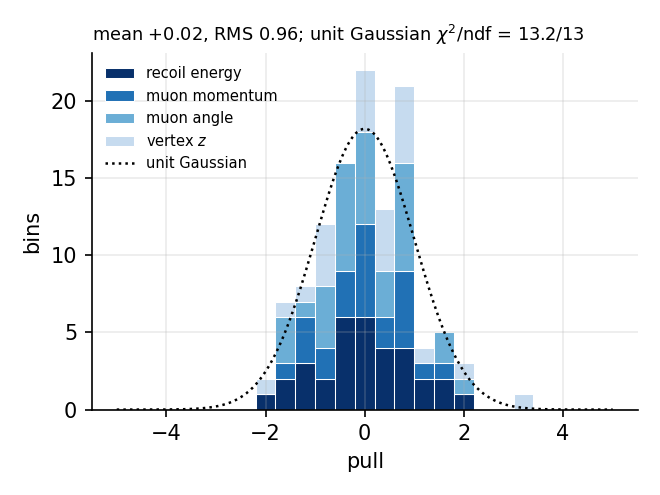}
\caption{Pulls of the marginal bin fractions of the recoil energy, the muon momentum and angle, and the vertex $z$, ensemble mean against the open simulation dataset on the test sample (Eq.~\ref{eq:pull}), with the open simulation dataset's statistical error, the surrogate's sampling error and the between-member spread in quadrature. Dotted: unit Gaussian.}
\label{fig:ens-pulls}
\end{figure}

The uncertainty is then validated against the open simulation dataset with pulls. For each bin $i$ of the marginal distributions of the recoil energy, the muon momentum and angle, and the vertex $z$ on the test sample, the pull, $p_i$, is defined as:
\begin{equation}
p_i = \frac{\bar{s}_i - d_i}{\sqrt{\sigma_{d,i}^2 + \sigma_{s,i}^2 + \sigma_{e,i}^2}},
\label{eq:pull}
\end{equation}
where $d_i$ is the open simulation dataset's bin fraction, $\sigma_{d,i}$ is its statistical error, $\bar{s}_i$ the mean of the four surrogate ensemble members, $\sigma_{s,i}$ is the surrogate's sampling error, and $\sigma_{e,i}$ is the surrogate ensemble spread. Figure~\ref{fig:ens-pulls} shows the pull distribution over \nEnsPullAllBins{} bins. Its mean close to zero and RMS of approximately one indicate that the residuals on the test sample are consistent with the combined statistical, sampling and ensemble uncertainties for the observables considered.

\section{Holdout Validation}\label{sec:holdout}
A central question is whether the surrogate can reliably predict the detector response to interactions it has not encountered during training. The universality of the surrogate rests on this: the regions of final-state phase space populated by different interaction models do not fully overlap, and a model applied to a new generator meets final states that the training simulation covered sparsely or not at all. We investigate this through a dedicated holdout study: excluding some events from the training sample and then testing the model performance when exposed to them. 2p2h interactions~\cite{Nieves2011}, in which the neutrino scatters off a correlated pair of nucleons, provide a particularly useful test of this capability. Typically characterized by low hadronic energy and pionless final states involving two outgoing nucleons, these interactions populate regions of final-state phase space that are poorly represented by other interaction modes, limiting the effectiveness of conventional reweighting.
We can demonstrate the lack-of-coverage using a classifier~\cite{Sugiyama2012,Cranmer2015} of 2p2h against all other modes which gives
the density of the other modes relative to 2p2h at each event:
\nCovTwoPLtEqual{} of the 2p2h events lie where the 2p2h-free
sample has fewer events than the 2p2h sample itself, and for
\nCovTwoPLtHundredth{} a reweighting would need weights above 100.

\begin{table}[tb]\centering\small
\begin{tabular}{lcc}
\toprule
closure AUC & full model & 2p2h holdout \\
 & 2p2h & 2p2h \\
\midrule
reconstructed only & \nAECloATwoPCIMarg{} & \nAECloBTwoPCIMarg{} \\
with truth features & \nAECloATwoPCICond{} & \nAECloBTwoPCICond{} \\
\bottomrule
\end{tabular}

\caption{Holdout summary. Columns: the full model and the 2p2h holdout model on the 2p2h test events. Closure AUCs on the seven event-level variables as in Table~\ref{tab:closure}; 0.5 means indistinguishable from the open simulation dataset. Both 2p2h columns are consistent with 0.5 within the statistical resolution of \nAECloATwoPRes{}.}
\label{tab:holdout}
\end{table}

\begin{figure*}[tb]\centering
\includegraphics[width=\textwidth]{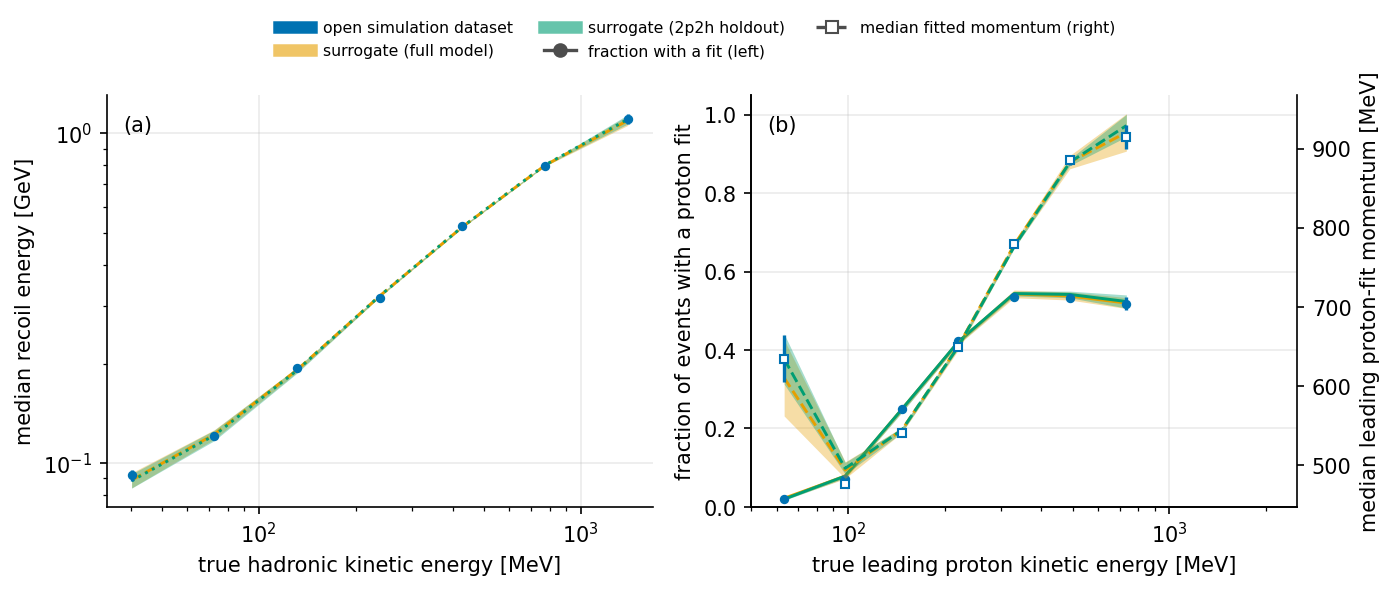}
\caption{As Figs.~\ref{fig:response} and~\ref{fig:hadron}, for the held-out 2p2h test events: (a) recoil energy and (b) proton reconstruction (2p2h final states contain no charged pions), for the open simulation dataset, the full model and the 2p2h holdout model. The bands are the surrogate's sampling spread, the statistical fluctuation of a sample of this size over eight regenerations with different random numbers and fixed weights, comparable to the open simulation dataset's error bars; for the full model the between-member spread is added in quadrature, and on this sample it is several times smaller than the sampling spread. The 2p2h holdout model has no ensemble, and its band is sampling only.}
\label{fig:hold-response}
\end{figure*}

\begin{figure*}[tb]\centering
\includegraphics[width=\textwidth]{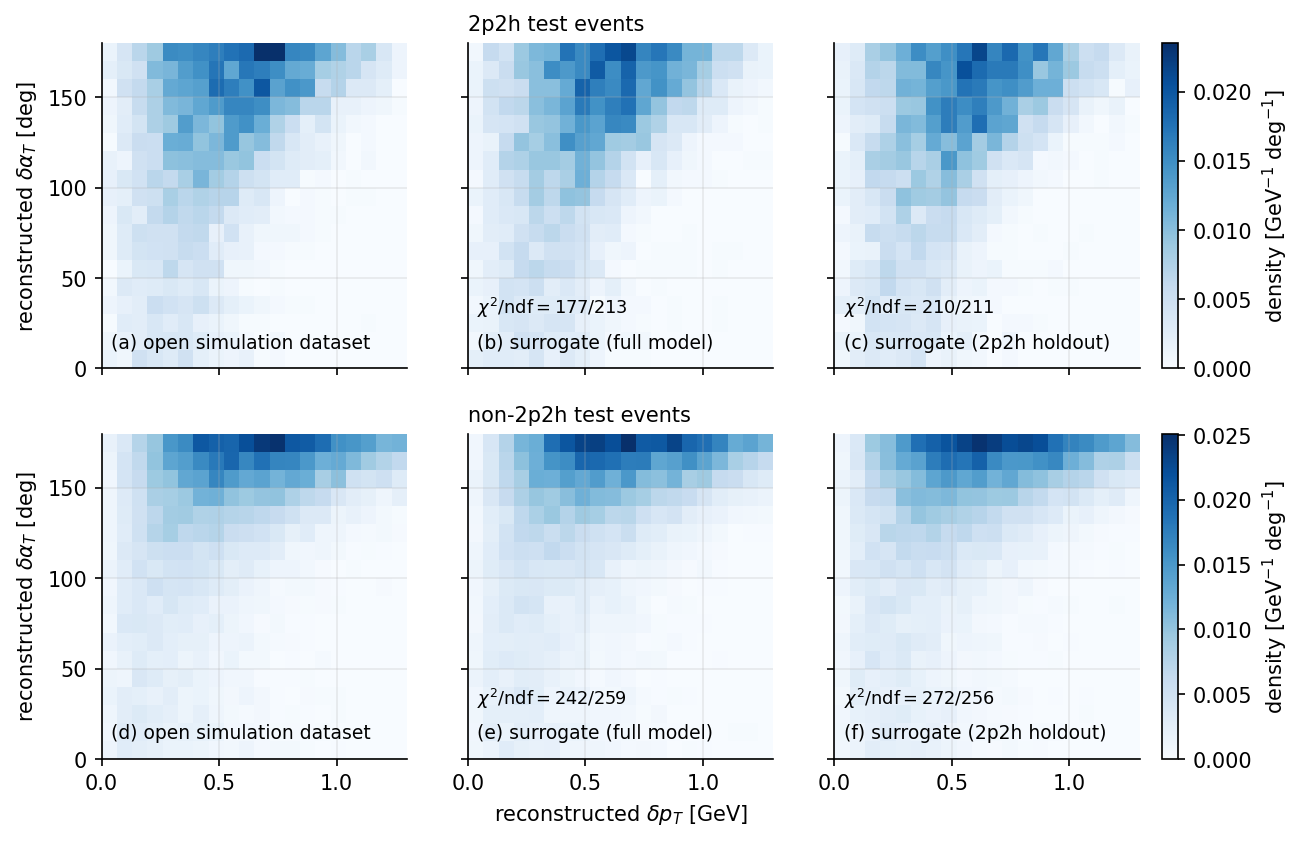}
\caption{Reconstructed transverse kinematic imbalance of the muon and the leading proton, $\delta\alpha_T$ against $\delta p_T$, for the 2p2h test events (top) and the non-2p2h $\nu_\mu$ charged-current test events (bottom): (a, d) the open simulation dataset, (b, e) the full model and (c, f) the 2p2h holdout model. Each surrogate panel quotes the two-sample reduced $\chi^2$ of its histogram against the open simulation dataset's over the bins with at least five events in both.}
\label{fig:hold-tki}
\end{figure*}

A second model, the 2p2h holdout model, is trained with all 2p2h events (337{,}617 events) removed, which we compare against the full model discussed up until now. Both are evaluated on the 2p2h test events compared with the original open data reconstruction of those events, and the 2p2h holdout model is also evaluated on the non-2p2h test events as a control. Recall, neither model receives the interaction mode as input: the test asks whether the detector response learned from the other modes transfers to 2p2h final states.

As shown in Table~\ref{tab:holdout}, the 2p2h holdout model achieves comparable closure to the full model on the 2p2h test events, while also maintaining agreement on the non-2p2h control sample.
Classifiers trained to distinguish the open simulation dataset from each model and the two models against each other indicate the same agreement: the 2p2h holdout model and the full model cannot be distinguished, with AUCs that both differ from the open simulation dataset by about 1\%. Figure~\ref{fig:hold-response} shows that the recoil energy and the proton reconstruction of the two models lie on top of each other and agree with the open simulation dataset within errors. Figure~\ref{fig:hold-tki} tests a lepton--hadron correlation, the transverse kinematic imbalance of the muon and the leading proton~\cite{Lu2016,Lu2018}. Both models reproduce the open simulation dataset within errors, for the distinctive 2p2h signature and for the non-2p2h events alike. The agreement suggests that the learned response transfers across interaction categories when the relevant final-state information is retained, even where the statistical limitations of reweighting make direct comparisons difficult.

\section{End-to-end MINERvA Analysis}\label{sec:analysis}

The MINERvA open data release~\cite{opendata} is accompanied by a tutorial analysis, ``MINERvA~101'', built on the collaboration's analysis toolkit. It measures the charged-current inclusive $\nu_\mu$ cross section on the scintillator tracker as a function of the muon transverse momentum $p_T$, using the Medium Energy forward-horn-current playlist~1A. The signal is a $\nu_\mu$ charged-current interaction with the vertex in the tracker fiducial volume, a muon angle below $20^\circ$ with respect to the beam and a muon longitudinal momentum above $1.5$~GeV/$c$. Events are selected with six reconstruction cuts: the vertex position along the beam, the vertex position in the transverse plane, the muon angle, a muon track matched in the MINOS near detector, low dead time upstream of the vertex, and a negative muon charge. The tutorial then takes the selected sample through background subtraction, unfolding, and efficiency and flux normalization to a differential cross section with systematic uncertainties.
Here, we take a different approach, using forward folding to bring theoretical predictions into reconstructed space for direct comparison with the data.
Since the goal is just to demonstrate the method, systematic uncertainties are not included.

\begin{figure}[tbp]
  \centering
  \includegraphics[width=\columnwidth]{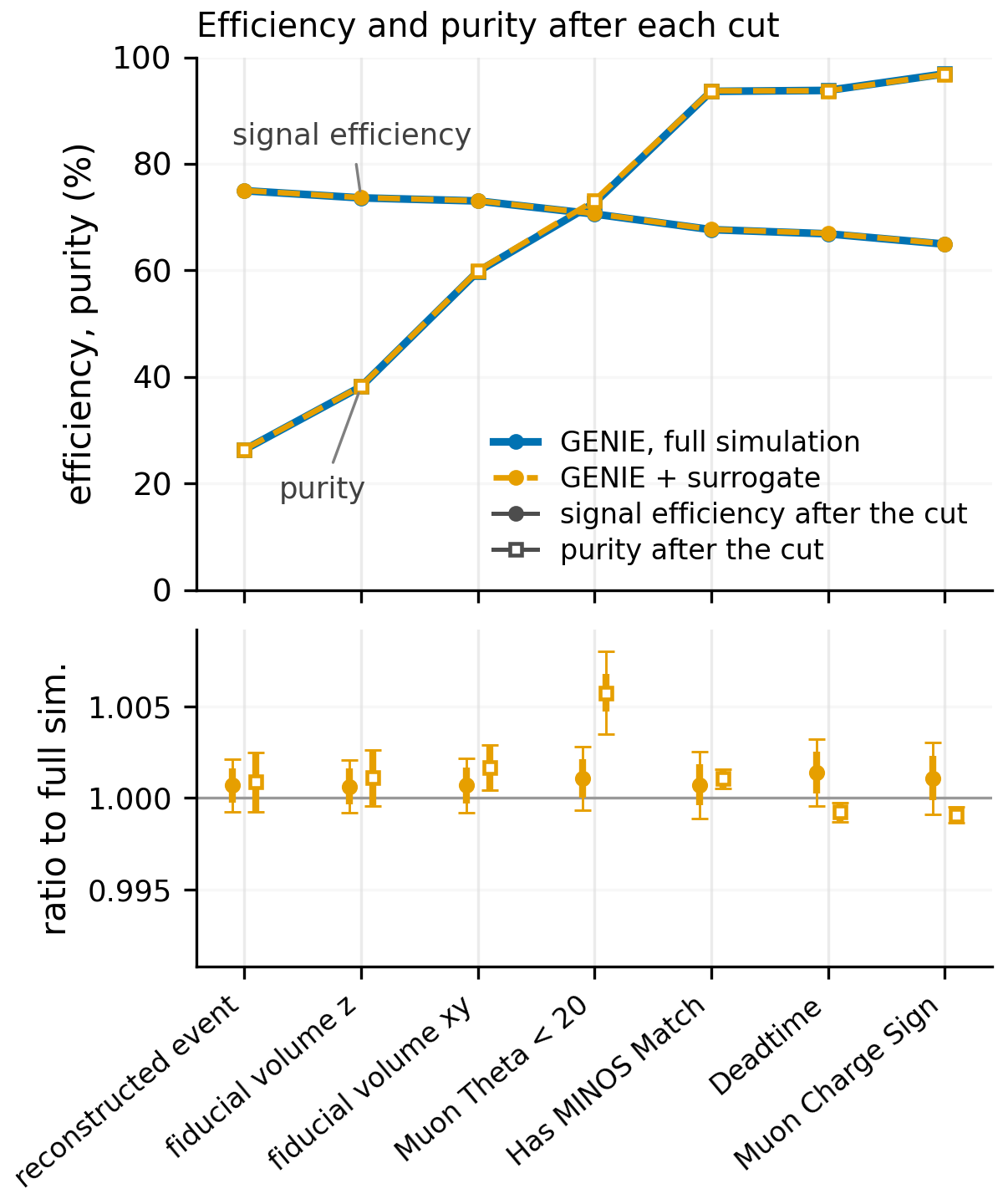}
  \caption{Signal efficiency (filled circles) and purity of the selected sample (open squares) after each cut of the MINERvA~101 selection, for the GENIE full simulation (all reconstructed events) and GENIE + surrogate (the same generated events) with their ratio as shown in bottom.}
  \label{fig:cutflow-eff-pur}
\end{figure}

We first perform a closure test using the official GENIE sample. Figure~\ref{fig:cutflow-eff-pur} follows the signal efficiency and the purity of the selected sample through the six cuts, for the full simulation and for the surrogate run on the same generated events.
The surrogate reproduces the signal efficiency of the full simulation within the estimated statistical and model uncertainties, and the purity within 2.5 standard deviations, at every stage, demonstrating its ability to capture the reconstruction and selection response for both signal and background events.

Having established closure for the CC-inclusive selection, a natural question follows: can the same learned detector response remain reliable as the physics selection becomes more demanding? We investigate this through two additional cutflow closure tests inspired by MINERvA's charged- and neutral-pion analyses~\cite{MINERvA:2022djk, MINERvA:2020anu}. Due to the samples used and the construction of the analyses, these selections are not reproduced in full and should not be compared with the published MINERvA results. In both cases, we retain the published signal definition and apply the subset of reconstruction cuts that can be evaluated using the surrogate output. Requirements that depend on observables not represented by the surrogate are omitted. These selections allow us to explore how well the learned detector response carries over to analyses with different reconstruction demands.
Four cuts of the MINERvA~101 selection are shared by both analyses: the two fiducial-volume cuts, the MINOS-matched muon track and the negative muon charge.

For the single charged pion analysis the signal is a $\nu_\mu$ charged-current interaction in the tracker fiducial volume with exactly one $\pi^+$ and no other meson in the final state.
The pion kinetic energy and the invariant mass of the hadronic system are restricted to $35 < T_\pi < 350$~MeV and $W < 1.4$~GeV, and the muon kinematics to $1.5 < p_\mu < 20$~GeV/$c$ and $\theta_\mu < 13^\circ$.
The selection requires at least two tracks from the vertex; exactly one hadron track with a successful pion fit, used here an an available proxy for pion-identification requirements of the published analysis; and the $W$, muon and pion ranges of the signal definition applied to reconstructed quantities.
The invariant mass is obtained with
\begin{equation}
    W^2\!=M_N^2+2M_N q_0 - 2 (E_\mu+q_0)(E_\mu-p_\mu\! \cos\theta_\mu)+m_\mu^2,
\end{equation}
where the calorimetric energy is defined as
\begin{equation}
    q_0=\sum_i^{\rm hadrons} \!E_i - (\text{number of baryons})\times M_N,
\end{equation}
with $M_N$ being the average of proton and neutron masses.

For the neutral pion analysis the signal is a $\nu_\mu$ charged-current interaction on carbon with at least one proton, at least one $\pi^0$, and no other meson or hyperon in the final state.
The leading proton momentum is required to be above $0.45$~GeV/$c$ and the muon kinematics are restricted to $1.5 < p_\mu < 20$~GeV/$c$ and $\theta_\mu < 25^\circ$.
The selection requires the muon kinematic ranges of the signal definition; that every non-muon track is identified as a proton by its score; and a leading proton momentum above $0.45$~GeV/$c$.
As Ref.~\cite{MINERvA:2020anu} does not quote the proton score threshold, we run the selection with two values, 0.5 and 0.8.
Table~\ref{tab:cutflow-final} lists the efficiency and purity after each of the three sets of selections. The agreement across these distinct selections is particularly encouraging: it suggests that the learned detector response is not confined to a single analysis, but can be carried across different final-state topologies and reconstruction requirements.

\begin{table}[t]
\centering
\begin{tabular}{llcc}
\hline\hline
Selection & Sample & Efficiency (\%) & Purity (\%) \\
\hline
CC inclusive & full sim. & $64.77 \pm 0.03$ & $97.08 \pm 0.01$ \\
Reference~\cite{opendata}     & surrogate       & $64.81 \pm 0.12$ & $96.99 \pm 0.04$ \\
\hline
Single $\pi^+$ & full sim. & $38.98 \pm 0.12$ & $31.13 \pm 0.11$ \\
Inspired by \cite{MINERvA:2022djk} & surrogate       & $38.21 \pm 0.25$ & $30.80 \pm 0.23$ \\
\hline
$p\pi^0$, $p$ score $> 0.5$ & full sim. & $12.70 \pm 0.08$ & $15.09 \pm 0.09$ \\
Inspired by \cite{MINERvA:2020anu} & surrogate       & $12.68 \pm 0.16$ & $15.00 \pm 0.18$ \\
\hline
$p\pi^0$, $p$ score $> 0.8$ & full sim. & $4.75 \pm 0.05$ & $14.72 \pm 0.14$ \\
Inspired by \cite{MINERvA:2020anu} & surrogate       & $4.67 \pm 0.10$ & $15.12 \pm 0.30$ \\
\hline\hline
\end{tabular}
\caption{Signal efficiency and purity after each of the three sets of selections, charged-current inclusive, single $\pi^+$ and $p\pi^0$, for the full simulation and for the surrogate.}
\label{tab:cutflow-final}
\end{table}

\begin{figure*}[tbp]
  \centering
  \includegraphics[width=\textwidth]{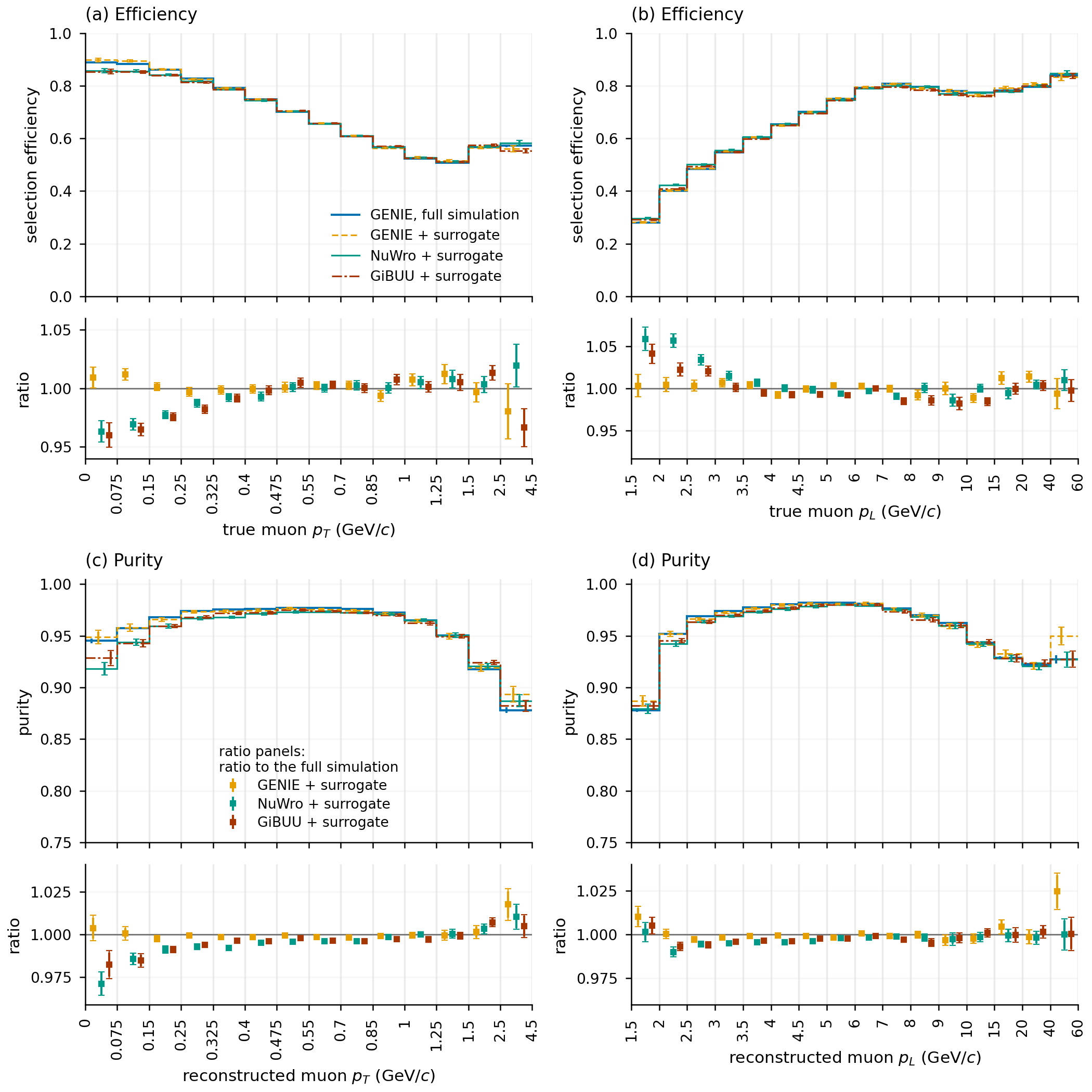}
  \caption{Selection efficiency for signal events against the true muon $p_T$ (a) and $p_L$ (b), and purity of the selected sample against the reconstructed muon $p_T$ (c) and $p_L$ (d), for the GENIE full simulation, GENIE + surrogate, NuWro + surrogate and GiBUU + surrogate, each with its ratio to the full simulation below; the full simulation is taken over all its reconstructed events. Thick bars are statistical; for the surrogate predictions the thin capped bars add the surrogate model uncertainty, in the fully correlated limit, in quadrature.}
  \label{fig:eff-pur}
\end{figure*}

We then apply the same surrogate model to NuWro~\cite{Golan2012} and GiBUU~\cite{Buss2012} samples generated with the MINERvA $\nu$-$e$ constrained flux~\cite{MINERvA:2019hhc}, with the interaction vertices distributed over the nuclides of the MINERvA detector according to its geometry model~\cite{Aliaga2014,opendata}. Figure~\ref{fig:eff-pur}(a) and (b) shows the selection efficiency in bins of the true muon momentum and Figure~\ref{fig:eff-pur}(c) and (d) shows the purity in bins of the reconstructed momentum, following the binning of Ref.~\cite{MINERvA:2021owq}, for the original GENIE-based simulation, as well as surrogate events from GENIE, NuWro, and GiBUU predictions. The closure test (full simulation vs. GENIE surrogate) passes for both $p_T$ and $p_L$ within the uncertainty of the surrogate model. Through the surrogate, NuWro and GiBUU give total efficiencies and purities similar to those of GENIE, with visible differences at low $p_L$ and low $p_T$.

\begin{figure*}[tbp]
  \centering
  \includegraphics[width=\textwidth]{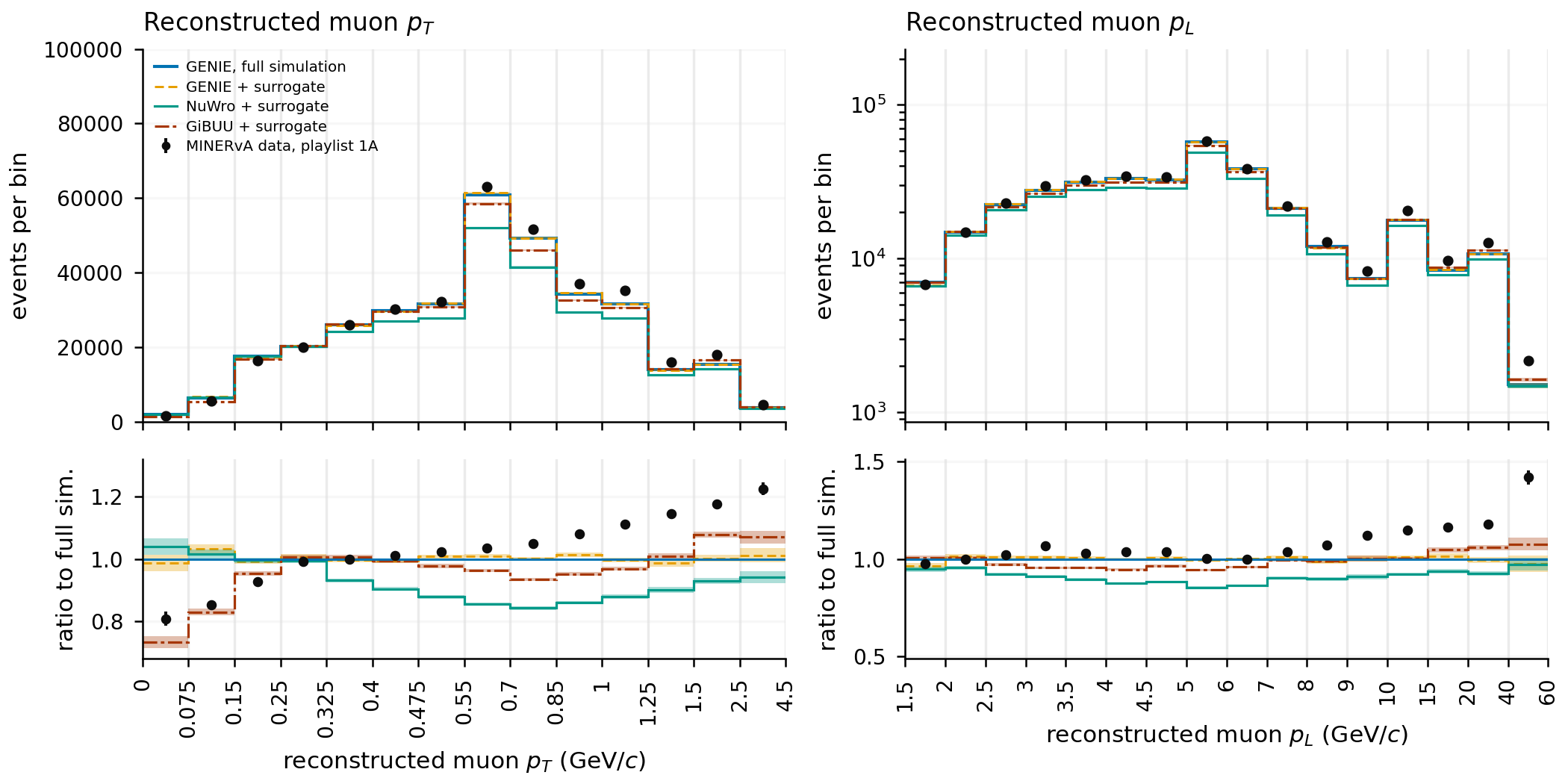}
  \caption{Selected events against the reconstructed muon $p_T$ (left) and $p_L$ (right) for the MINERvA playlist-1A data, the GENIE full simulation, GENIE + surrogate, NuWro + surrogate and GiBUU + surrogate, each with its ratio to the GENIE full simulation. The data are raw counts with statistical errors. The bands are the statistical uncertainty of the predictions and, for the surrogate predictions, the surrogate model uncertainty in the fully correlated limit added in quadrature.}
  \label{fig:muon-1d}
\end{figure*}

Figure~\ref{fig:muon-1d} demonstrates the central goal of this work: a direct comparison between experimental data and different neutrino interaction models in reconstructed space, normalized by exposure (protons on target) to show total cross section differences. The closure test holds here as well: through the surrogate model, we reproduce the selected-event distributions of the full simulation in $p_T$ and $p_L$ within the surrogate model uncertainty.

The two other generators, passed through the same surrogate, give predictions that are clearly different from the closure test and from each other, well outside the model uncertainties.
We have checked that the different trends among generators in Fig.~\ref{fig:muon-1d} are also present in the true muon $p_T$ and $p_L$ distributions for the same signal definition.

This comparison required none of the generator-dependent steps of a cross-section extraction: the data are the raw selected events binned in simple observables, and each generator is folded through the same detector response, provided by the surrogate, and compared with them directly. Background subtraction, the unfolding of the migration between true and reconstructed bins with its regularization, and the efficiency correction are not needed; in a cross-section extraction each of these is built from the simulation of a single generator and carries that generator's model into the measurement. The rest of the uncertainty sources remain, including the statistical uncertainty of the data, the uncertainties of the generator predictions, the uncertainties of the detector model that the surrogate learned from the simulation, which enter the forward-folded prediction as they would enter an efficiency correction or a migration matrix, and the surrogate's own model uncertainty. The present demonstration includes the latter, estimated to be below one percent per bin, but does not yet incorporate the full experimental and theoretical systematic uncertainties.

\section{Conclusions}\label{sec:conclusion}

Making experimental data publicly available is only the first step in making it scientifically reusable. The
ability to reinterpret those measurements has remained closely tied to the simulation and reconstruction machinery of
the original experiments. In this work, we present the design, performance, and validation of a surrogate model for
the MINERvA detector simulation and reconstruction, trained entirely on the MINERvA open simulation dataset. We have
demonstrated that such a surrogate model can help bridge that gap. By learning the relationship between
generator-level final states and reconstructed observables, rather than relying on generator-specific descriptions of
the underlying interaction, the surrogate provides a common detector-response framework through which alternative
theoretical predictions can be explored.

The strength of this approach lies not only in its ability to reproduce the original simulation, but in how far that
learned response can be carried beyond it. The close agreement across reconstructed observables, supported by
classifier-based closure tests, establishes the fidelity of the surrogate within the tested distributions, and within
the uncertainties reported by the surrogate model through its ensemble. More importantly, the 2p2h holdout study
offers a particularly encouraging demonstration of transferability: even without encountering this interaction
category during training, the surrogate reproduces its reconstructed response with fidelity comparable to that of a
model trained on the full sample. This opens a route to investigating interaction models that are poorly represented
in existing simulations, where conventional reweighting becomes statistically limited. We put this capability to use
in a MINERvA charged-current inclusive analysis, bringing predictions from GENIE, NuWro, and GiBUU into reconstructed
space for direct comparison with experimental data. The resulting generator-dependent differences remain visible after
the detector response is applied, demonstrating the value of forward folding as a means of confronting alternative
interaction models with measurements.

Although demonstrated here for a single MINERvA running configuration, the approach offers a foundation for extending
these comparisons to other playlists, beam modes~\cite{Adamson2016,Valencia2019,Zazueta2023}, and systematic
variations~\cite{Aliaga2016}. Such a surrogate model can also be adapted to other open simulation
releases~\cite{Cerati2024}, following the same backbone with a different encoding of the inputs and decoding of the
outputs, to be discussed in future publications. Paired with the rapidly growing capabilities of agentic AI
systems~\cite{Yao2023,Diefenbacher2025,GendreauDistler2025}, such a surrogate lets a researcher rapidly translate a
hypothesis and a novel model into a measurement. More broadly, this work points towards a different role for open
experimental data: not simply as an archive of completed measurements, but as a starting point for new questions. By
making the connection between theoretical predictions and detector observables more accessible, surrogate models can
help turn that possibility into a practical avenue for independent scientific investigation.

The surrogate model presented here, with its training code and trained weights, is openly available at \url{https://github.com/fnal-nd-aiml/XSuM-MINERvA}.

\section{Acknowledgments}\label{sec:acknowledgements}

This work was produced by Fermi Forward Discovery Group, LLC under Contract No. 89243024CSC000002 with the U.S. Department of Energy, Office of Science, Office of High Energy Physics, with the support of the Lab Directed Research and Development program. Publisher acknowledges the U.S. Government license to provide public access under the DOE Public Access Plan. The work of D.L. is supported by U.S. Department of Energy, Office of Science, Office of High Energy Physics under Award Number DE-SC0008475.
The authors thank the MINERvA Collaboration for making their data, simulated data, and analysis tools available to the community. We thank Heidi Schellman for discussion. The authors made use of agentic AI in tools in the development of the software presented here, for visualizing results, and for editorial assistance in revising the paper. The authors have carefully verified all work and stand by what is presented.

\end{document}